\documentclass[aps,showpacs,showkeys,prd,preprintnumbers,nofootinbib]{revtex4}

\usepackage{color}
\usepackage{graphicx}
\usepackage{bbm}

\newcommand{\nc}{\newcommand}
\nc{\beq}{\begin{equation}}  \nc{\eeq}{\end{equation}}
\nc{\bea}{\begin{eqnarray}}  \nc{\eea}{\end{eqnarray}}
\nc{\bpm}{\begin{pmatrix}}   \nc{\epm}{\end{pmatrix}}
\def\xx{{\bf x}}
\def\yy{{\bf y}}
\def\CC{{\bf C}}
\def\pp{{\bf p}}
\def\kk{{\bf k}}
\def\KK{{\bf K}}
\def\then{{\quad\Rightarrow\quad}}
\def\ket#1{\left| #1 \right\rangle}

\def\bcal{{\cal B}}

\def\gcal{{\cal G}}
\def\lcal{{\cal L}}
\def\mcal{{\cal M}}
\def\ncal{{\cal N}}
\def\ocal{{\cal O}}
\def\ucal{{\cal U}}
\def\zcal{{\cal Z}}
\def\gesim{\,{\raise-3pt\hbox{$\sim$}}\!\!\!\!\!{\raise2pt\hbox{$>$}}\,}
\def\lesim{\,{\raise-3pt\hbox{$\sim$}}\!\!\!\!\!{\raise2pt\hbox{$<$}}\,}
\def\su#1{{SU(#1)}}

\def\tev{\hbox{TeV}}
\def\up#1{^{\left( #1 \right) }}
\def\vevof#1{\left\langle#1\right\rangle}
\def\rhs{right hand side\ }
\def\lhs{left hand side\ }
\def\tr{{\rm Tr}}
\def\deriva#1#2#3{\left(\frac{\partial #1}{\partial #2}\right)_{#3}}
\def\half{\frac12}
\def\inv#1{\frac1{#1}}
\def\qwe{\delta}
\def\gsim{\gesim}
\def\lsim{\lesim}

\nc\ssm{{\rm SM}}
\nc\snp{{\rm NP}}
\nc{\du}{d_\ucal}
\nc{\dsm}{d_\ssm}
\nc{\bz}{{\bcal\zcal}}
\nc{\dbz}{{d_{\bcal \zcal}}}
\nc{\lu}{\Lambda_\ucal}
\nc{\lun}{\Lambda_{\not\ucal}}
\nc{\ou}{\ocal_\ucal}
\nc{\osm}{\ocal_{SM}}
\nc{\obz}{\ocal_{\bcal \zcal}}
\nc{\mpl}{M_{Pl}}
\nc{\tbz}{T_{\bcal\zcal}}
\nc{\mc}{M_\ucal}
\nc{\cu}{C_\ucal}
\nc{\tf}{T_f}
\nc{\tfu}{T_{\ucal\hbox{\tiny-}f}}
\nc{\tfbz}{T_{\bz\hbox{\tiny-}f}}
\nc{\tsm}{T_\ssm}
\nc{\tnp}{T_\snp}
\nc{\tew}{T_{\rm EW}}
\nc{\tu}{T_\ucal}
\nc{\gsm}{g_\ssm}
\nc{\gu}{g_{\ucal}}
\nc{\gbz}{g_\bz}
\nc{\gir}{g_{\rm IR}}
\nc{\gnp}{g_\snp}
\nc{\gtot}{g_{\rm tot}}
\nc{\dg}{f}
\nc{\rsm}{\rho_{\rm SM}}
\nc{\ru}{\rho_\ucal}
\nc{\rbz}{\rho_\bz}
\nc{\rnp}{\rho_{\rm NP}}
\nc{\rtot}{\rho_{\rm tot}}
\nc{\co}{{\rm const.}}
\nc{\Pu}{P_\ucal}
\nc{\lm}{g}

\begin{document}

\preprint{IFT-08-10 \cr UCRHEP-T455}

\title{Cosmology with Unparticles}

\author{Bohdan GRZADKOWSKI}
\email{bohdan.grzadkowski@fuw.edu.pl}
\affiliation{Institute of Theoretical Physics,  University of Warsaw,
Ho\.za 69, PL-00-681 Warsaw, Poland}

\author{Jos\'e WUDKA}
\email{jose.wudka@ucr.edu}
\affiliation{Department of Physics and Astronomy, University of California,
Riverside CA 92521-0413, USA\\
and\\
Departamento de F\'\i sica Te\'orica y del Cosmos\\
Universidad de Granada E-18071, Granada, Spain}

\begin{abstract}

We discuss cosmological consequences of the existence of physics
beyond the standard model that exhibits Banks-Zaks and unparticle
behavior in the UV and IR respectively. We first derive the
equation of state for unparticles and use it to obtain the temperature
dependence of the corresponding energy and entropy densities.  We then
formulate the Boltzmann 
and Kubo equations
for both the unparticles and the
Banks-Zaks particles, and use these results to determine the
equilibrium conditions between the standard model and the new
physics. We conclude by obtaining the constraints on the effective
number of degrees of freedom of unparticles imposed by Big-Bang
nucleosynthesis.

\end{abstract}


\maketitle

\section{Introduction}

Recently Georgi~\cite{Georgi:2007ek, Georgi:2007si}\footnote{A similar idea
was discussed also in \cite{van der Bij:2006pg}.} raised the
interesting possibility that physics beyond the Standard Model (SM)
may contain a sector that is conformally invariant in the IR region
(guaranteed by a zero of the beta function), 
and classically
scale-invariant in the UV; we refer to these as the
unparticle ($\ucal$) and Banks-Zaks ($\bz$) phases, respectively.  The
transition region between the two phases is characterized by the scale
of dimensional transmutation $\lu$ 
A specific realization of this
idea can be found in \cite{Banks:1981nn}; following this reference
we will assume that the new sector is described as an
asymptotically free gauge theory in the $\bz$ phase.

This novel idea has received substantial attention within the
high-energy community, 
mainly in connection with the phenomenology of such models.
Here we discuss some fundamental issues in the
evolution of the Universe in the presence of this type of new
physics (though 
studies of the cosmological consequences of the proposal have
appeared in the literature
~\cite{Davoudiasl:2007jr}-\cite{Kikuchi:2007az},
these publications ignore
several essential aspects which are discussed below).
In sec.~\ref{sec:stateq} we
derive an approximate equation of state for the NP sector. Then,
in sec.~\ref{sm-np} we
use this together with the expected SM-NP 
interactions~\cite{Georgi:2007ek, Georgi:2007si} to
determine the conditions under which the SM and NP sectors
were in equilibrium. 
In sec.~\ref{bbn}, using the experimental 
constraints derived from Big-Bang Nucleosynthesis (BBN) we obtain
non-trivial  bounds on the parameters of the theory.
The Appendices A and B are devoted to presentation of two
alternative derivation of the Boltzmann equation.


\section{Thermodynamics of unparticles}
\label{sec:stateq}
In order to understand the thermodynamic behavior 
of the new sector~\footnote{The
thermodynamics of conformal theories has been studied
extensively~\cite{cft.temp}, but these results have been apparently
ignored where unparticles are concerned.} we
use the expression for the trace anomaly of the energy momentum tensor
of a gauge theory where all the renormalized masses
vanish~\cite{Collins:1976yq}:
\beq
\theta_\mu^\mu = \frac\beta{2\lm} N \left[F^{\mu\nu}_a F_{a \; \mu\nu}
\right]\,,
\label{eq:trace}
\eeq
where $ \beta $ denotes the beta function for the coupling $\lm$ and
$N$ stands for the normal product.

The basic assumption for the unparticle phase is that the $\beta $
function has a non-trivial IR fixed point at $ \lm = \lm_\star 
\not=0
$. 
Modeling the unparticle sector by a
 gauge theory, we assume that for
low temperatures~\footnote{The cases where $ \beta $
has a higher-order zero at $ \lm_\star$ can be treated similarly.}
\beq
\beta = a ( \lm - \lm_\star) , \quad a>0 \,,
\eeq
in which case the  running coupling reads
\beq
\lm(\mu) = \lm_\star + u \mu^a; \qquad
\beta[\lm(\mu)] = a u \mu^a \,,
\label{bet}
\eeq
where $u$ is an integration constant and $\mu$ is the renormalization
scale.

We look for the lowest-order corrections to the conformal limit (where
$\theta^\mu_\mu =0 $) when the system is in thermal equilibrium at
temperature $T$, is isotropic and homogeneous, and does not have any
net conserved charge. Since $ \beta $ vanishes in the conformal limit,
in (\ref{eq:trace}) we can take $ \vevof{N \left[F^{\mu\nu}_a F_{a \;
\mu\nu} \right]} $ equal to its conformal value (we denote the thermal
average by $ \vevof{\cdots} $); taking the renormalization scale $\mu
= T$ we then expect
\beq
\vevof{N \left[F^{\mu\nu}_a F_{a \; \mu\nu} \right]} = b T^{4+\gamma }\,,
\label{FFav}
\eeq 
where $ \gamma $ is the anomalous dimension of the operator.  Using $
\vevof{\theta_\mu^\mu} = \ru - 3 \Pu $, where $ \ru $ and $\Pu$ denote
the energy density and pressure of the unparticle phase, together with
(\ref{bet}) and (\ref{FFav}) then gives
\beq
\ru - 3 \Pu = A T^{4+\qwe}; \quad
\left( A \equiv \frac{a u b}{2 \lm_\star}, ~\qwe\equiv a+ \gamma \right)\,,
\label{eq:treq}
\eeq
where we took $ \mu = T $.

Combining (\ref{eq:treq}) with the thermodynamic relation $ d(\rho V)
+ P \; dV = T \; d(s V) $ ($s$ is the entropy density),
when $ \rho$ and $P$ are functions of $T$
only\footnote{A consequence of having assumed the absence of net
charges.}, and integrating, we find,
\bea
\ru     &=&   \sigma T^4   + A\left(1+\frac{3}{\qwe}\right) T^{4+\qwe} \cr 
\Pu     &=&   \frac13 \sigma T^4 + \left(\frac{A}{\qwe}\right)    T^{4+\qwe} \cr 
s_\ucal &=&   \frac43 \sigma T^3 + A\left(1+\frac{4}{\qwe}\right) T^{3+\qwe} 
\label{u.eos}
\eea
where $ \sigma $ is an integration constant and we assumed $ \qwe \not= 0$. 

It is worth noticing 
that the terms $ \propto A $ correspond to deviations from
the standard relativistic relation 
$V \propto T^{-3}$.
The behavior at low temperatures depends on the sign of $\qwe$,
we will assume $ \qwe > 0 $. Then
\beq
3\Pu =  \ru \left[1 - B \ru^{\qwe/4} \right]; \quad
B = \frac{A }{\sigma^{1+\qwe/4  }}
\eeq
exhibiting the lowest-order corrections to the often-used expression $
P = w \rho $, $ w = $const. 
This effect might be of interest
in the discussion of the possible dark-energy effects contained
in this model, but will not be discussed here.

Elucidating the cosmological effects 
of the modified equation of state
(\ref{u.eos}) lies beyond the 
scope of the present paper, we merely remark
that the NP  increases the coefficient of the $T^4$ term in $ \rho $
and induces $O(T^\qwe)$ corrections; e.g.
in the radiation-dominated era the scale parameter behaves
as $(1+c T^\qwe)^{1/3}/T $ ($c=$const.).

In general we expect $ A \propto \Lambda_\ucal^{-\qwe} $ since $ \lu $
is the scale associated with broken scale invariance; then the energy
density for the new sector in the unparticle phase equals
\beq
\ru = \frac{3}{\pi^2} T^4 \left[\gir + \left( \frac{T}{\lu} \right)^\qwe
\dg
\right]; \qquad T \ll \Lambda_\ucal
\label{eq:rhoir}
\eeq
where we replaced $ \sigma = 3 \gir/\pi^2 $ (hereafter we use 
the normalization from Maxwell-Boltzmann statistics)
and $\gir$,
the effective number of relativistic
degrees of freedom (RDF), will be estimated below.

In the $ \bz $ phase we assume the theory is asymptotically free so
that, up to logarithmic corrections,
\beq
\rbz = \frac{3}{\pi^2} \gbz T^4 ; \qquad T \gg \Lambda_\ucal 
\eeq
where $ \gbz $ denotes the RDF in this phase.

For intermediate temperatures the explicit form of the
thermodynamic functions requires a complete non-perturbative
calculation and the choice of a specific model; 
fortunately we will not need to consider the detailed 
behavior of the system. Given that $ \rho \propto T^4 $ 
in both  the IR and UV regions, for our purposes it will be sufficient
to use the interpolation
\beq
\rnp \equiv \frac{3}{\pi^2} \gnp T^4; \quad 
\gnp = \gbz \theta(T - \lu) + \gu \theta(\lu - T) 
\label{rho}
\eeq
where $\gu=[\gir + \left(T/\lu \right)^\qwe \dg]$ while NP stands for
`new physics'; $\gnp$  will be continuous at $T=\lu$ when $ \dg
=\gbz-\gir$, which we now assume.  It is worth noting that a mass
distribution of unparticles with the spectral density $ \propto
(\mu^2)^{(\du-2)}$~\cite{Georgi:2007ek} 
generates the term $ \propto \dg $ in (\ref{rho})
with $\delta=2(\du-1)$, assuming that the contributions with $\mu > T$
decouple. We emphasize that (\ref{rho}) will be used only as a
rough but convenient approximation that reproduces the expected
behavior at low and high temperatures.
In cases of interest we expect $ \gir \sim \gbz \gg f $ so that the
terms $ \propto T^\delta $ are subdominant. 

Estimating $ \gir $ directly form the model Lagrangian
is a non-trivial exercise, due to the expected strong-coupling nature 
of the theory in the infrared. Using, however, the
AdS-CFT correspondence \cite{Gubser:1999vj} we find
\beq
\gir =  \frac{\pi^5}8 (L \mpl)^2
\eeq
where
$L$ denotes the AdS radius of curvature and $\mpl$ is the 
Planck mass. Given that $L$ is expected  \cite{Gubser:1999vj} to be 
significantly smaller than $ 1/\mpl $, it is 
justified to expect that
\beq
\gir \gesim \ocal(100)
\eeq
In the following we will use this as our estimate for the RDF
in the unparticle phase.

In order to estimate $ \gbz$ one must specify the details of the 
non-Abelian theory in the ultraviolet regime. For the models
considered in \cite{Banks:1981nn} we find
\beq
\gbz \sim 100
\label{eq:gbz.estimate}
\eeq
This result is based on a model
for which the couping constant stays within the perturbative regime
throughout its evolution. There is also non-perturbative lattice evidence
\cite{Svetitsky:2009pz} that 
gauge theories exhibiting an infrared fixed point
obey (\ref{eq:gbz.estimate}). In the following we will adopt this estimate.

The energy density 
$\ru $ was also discussed in \cite{Chen:2007qc}, however the expression
presented in this reference
agrees with (\ref{eq:rhoir}) only when $ \gir=0 $ and
therefore does not include the leading low-temperature behavior of the
theory.

\section{SM-NP interactions and equilibrium}
\label{sm-np}

The presence of a NP sector of the type considered here can have
important cosmological consequences since, even when weakly coupled
to the SM, its energy density will affect 
the expansion rate of the universe; this 
can then be used to obtain useful limits on the effective number of
degrees of freedom $ \gnp $. This calculation requires
a determination of the relationship between the temperature 
of the NP and SM sectors to which we now turn.

The interactions we will consider have the generic
form 
\beq
\lcal_{\rm int} = \epsilon \ocal_\ssm \ocal_\snp
\label{eq:Lint}
\eeq
where the first term is a
gauge invariant operator composed of SM fields (possible Lorentz
indices have been suppressed), while the second operator is either 
composed of
$\bz$  fields or is an unparticle operator, depending
on the relevant phase of the NP sector. The coupling $ \epsilon$
in general has dimensions and is assumed to be small. For the
specific calculations presented  below we will assume
for simplicity that $ \ocal_{\ssm,\snp} $ are both scalar operators.

Leading interactions
involve SM operators that can generate 2 particle states since states with
higher particle number will be phase-space suppressed. From such
interactions we obtain the NP$\leftrightarrow$SM reaction 
rate $ \Gamma$, which will be precisely defined below.
The two sectors will then
be in equilibrium whenever $ \Gamma \gsim H $,
where $H$ denotes the Hubble parameter \cite{Kolb}, 
and decouple at the transition temperature
$\tf$:
\beq
T=\tf:~\Gamma \simeq H; \quad 
H^2 = \frac{8 \pi}{3 \mpl^2} \rtot;
\label{fz}
\eeq
where
\bea
\rtot &=& \rsm + \rnp \cr
 &=& \frac3{\pi^2} \left( \gsm \tsm^4 + \gnp \tnp^4 \right)
\label{eq:tot.en.de}
\eea
We denote by $\tsm$ and $\tnp $
the temperatures for the SM and NP sectors which can be
different when these sectors are not in equilibrium

The approach to equilibrium can be described using
either  the Kubo formalism (appendix \ref{sec:app.kubo})
or a suitable extension of the
Boltzmann equation formalism 
(appendix \ref{sec:app.boltzmann}). It follows form the
expressions derived in the appendices that the conditions
near equilibrium are determined by the equation
\beq
\dot\vartheta + 4 H \vartheta = - \Gamma \vartheta; \;\;\;\;
\vartheta = \tnp - \tsm 
\label{eq:kubo.result}
\eeq
where, using the Kubo formalism,
\bea
\Gamma &=& \frac{\pi^2}{12 T^4} \left( \inv\gsm + \inv
\gnp \right) \epsilon^2 \Re \left\{
\int_0^\beta ds \int_0^\infty dt \int d^3\xx\;
\vevof{\ocal_\ssm(-i s , \xx) \dot\ocal_\ssm(t,{\bf0})}
\vevof{\ocal_\snp(-i s , \xx) \dot\ocal_\snp(t,{\bf0})} \right\}
\label{eq:kub}
\eea
The Boltzmann equation (BE) calculation
also yields (\ref{eq:kubo.result})
with the rate given by
\beq
\Gamma = \frac{\pi^2}{12 T^3}\left( \inv{\gsm} + \inv{\gnp} \right)
\inv{2 T}  \sum_{X',X}
\int d\Phi_\snp d\Phi_\ssm \beta (E_\ssm - E_\ssm')^2 e^{ - \beta E_\ssm}
\left| \mcal\right|^2 (2\pi)^4 \delta (K_\ssm - K_\snp)
\label{eq:bol}
\eeq
where
$ \mcal $ is the matrix element (with
no spin averaging) derived form the
SM-NP interaction Lagrangian~\footnote{The SM-SM and NP-NP 
interactions are not included because of our assumption that 
each sector is in equilibrium: these processes are much
faster than the ones generated by (\ref{eq:Lint}) and insure
that each sector has a well-defined temperature at all times.},
$ E_\ssm $ and $ E_\ssm'$ denote the
initial and final energies of the Standard Model particles in
the reaction, and $ K_{\ssm,\snp}$ the total  4-momenta
of each sector for the reaction; we have also
assumed the Boltzmann approximation 
(neglecting Pauli blocking or Bose-Einstein enhancement)
and denoted by
$ d\Phi_{\rm SM,NP}$ the appropriate
phase-space measures (without any spin factors).
In particular, for the unparticle
phase we use~\cite{Georgi:2007ek}
\beq
 d\Phi_\ucal = A_{\du}\epsilon(q^0)\; \theta(q^2)\; (q^2)^{\du-2} 
\frac{d^4 q}{(2\pi)^4}
\label{ps_u}
\eeq
where $ A_n = (4 \pi)^{3-2n}/[ 2 \Gamma(n) \Gamma(n-1)] $.
We show in  appendix \ref{sec:app.boltzmann}
that (\ref{eq:kub}) and (\ref{eq:bol}) are, in fact, equal.

The solutions to (\ref{eq:kubo.result}) yields  $\rho \propto R^{-4}$ in the
absence of the collision term 
(proportional to $ \Gamma $), 
as expected for a scale
invariant theory.
It is also important to note that, in contrast to other authors
(\cite{Davoudiasl:2007jr}-\cite{Lewis:2007ss}),
(\ref{eq:bol}) contains an
unparticle-decay term (see appendix \ref{sec:app.boltzmann}), as
we find the arguments (based on the
deconstruction picture~\cite{Stephanov:2007ry}) for neglecting
these contributions unjustified\footnote{
 (\ref{eq:bol}) 
gives the same result within the
unparticle scenario or the deconstruction approach;
in the latter case the vanishingly small
coupling constant of the deconstructed field is compensated by the
large number of particles of the same invariant mass
in the initial state. Unparticle decay
was discussed recently in \cite{Rajaraman:2008bc}.}.

The detailed calculation of $ \Gamma $ requires a specific 
form of the interaction $ \ocal_\ssm \ocal_\snp $ (see above for a
specific example). 
However for the purposes of the remaining calculations
only the basic properties of $ \Gamma$, such as its
 dependence on $T$ and the relevant RDF will be needed.
These properties can be obtained  using dimensional
analysis: if the dimensions of the operators are, respectively
$ d_\ssm $ and $ d_\snp $ and if the number of degrees
of freedom involved in this interaction are $ \gsm' $ and $ \gnp'$, 
then, including a phase-space factor we find
\beq
\Gamma \sim \frac{\epsilon^2 \lambda \gtot}{(4\pi)^{n_\ssm + n_\snp - 1}}
 T^{2 d_\ssm + 2 d_\snp - 7}; \quad
\lambda \equiv \frac{\gsm'}\gsm \frac{\gnp'}\gnp\,,
\label{gam}
\eeq
where $n_\ssm$ and $n_\snp$ denote numbers of SM and NP fields 
in the corresponding operators; in the unparticle phase we take
$ \gnp' = \du $ and
$ n_\snp =2( \du - 1 )$, where $ \du $ denotes
the dimension of $ \ocal_\ucal $.

The value of $ \lambda$ depends on the details
of the model. 
Above the Higgs ($\phi$) mass $m_\phi$ 
(we assume $m_\phi \sim v \equiv \vevof\phi$)
the most important operator is $\ocal_{\rm SM} = \phi^\dagger\phi$;
in this case $ \gsm'=4$, so 
$ \lambda \sim (4/\gsm) \cdot (\gnp'/\gnp)$. 
Below $m_\phi$ there are many dimension 4 SM operators relevant for the
SM-NP equilibration, {\it e.g.}
$\bar\ell \phi e $ (containing an extra suppression by 
the factor $\sim v/\mc$; $\ell, e$ denote a lepton isodoublet
and isosinglet respectively), or 
$B_{\mu\nu} B^{\mu\nu}$ (where $B$ is the hypercharge gauge field),
in this case we expect $ \gsm' \sim \gsm $, so that $ \lambda \sim \gnp'/\gnp$.


\subsection{{\it The Banks-Zaks phase.}}
We will assume that the $\bz$ sector corresponds to an $SU(n_c)$
Yang-Mills theory with $n_f$ vector-like massless fermions in 
the fundamental representation (denoted by
$ q_\bz$).  Assuming that $ \lu > v $,
the leading SM$\leftrightarrow$NP interaction is of the form
\beq
\lcal = \inv\mc \left(\phi^\dagger\phi \right)\left(\bar q_{\bz}q_{\bz}\right)
\label{lbzsm}
\eeq
where we assume that all flavors in the $\bz$ sector 
couple with the same strength.
In this case ($\epsilon =  1/\mc$)
\beq
\Gamma_\bz \simeq 
 \frac{\lambda \gtot}{(4\pi)^3 \mc^2 } T^3
\label{bzgam}
\eeq Denoting by $\tfbz$ the solution to (\ref{fz}) when $ \Gamma$ is given
by (\ref{bzgam}), and imposing also the consistency conditions $ \mc >
\tfbz > \lu $, we obtain ( $\gtot $ is evaluated at $ \tfbz $)
\beq
1 > \frac{\tfbz}{\mc} = \frac{\sqrt{(8\pi)^5 \gtot}}{\lambda \gtot}\;
\frac{\mc}{\mpl} > \frac{\lu}{\mc}
\label{tfbz}
\eeq


\subsection{{\it The unparticle phase.}}
In this case we will consider only
interactions of the form~\cite{Georgi:2007ek}
($k=\dsm+\dbz-4$)
\beq
\lcal = \frac{\lu^{\dbz-\du}}{\mc^k} \ocal_\ssm \ocal_\ucal
\label{lusm}
\eeq
Using (\ref{gam}) we obtain (here we use $ n_\snp = 2(\du-1) $)
\beq
\Gamma_\ucal \sim
\frac{\lambda \gtot \lu}{(4\pi)^{n_\ssm + 2\du - 3}}
 \left( \frac\lu\mc \right)^{2k}
\left( \frac T\lu \right)^{2 \dsm + 2 \du - 7}
\label{ugam}
\eeq
Denoting by $\tfu$ the solution to (\ref{fz}) when $ \Gamma$ is given
by (\ref{ugam}), and imposing also the consistency condition $ \lu >
\tfu $, we obtain (here $ \gtot $ is evaluated at $ \tfu $)
\beq
\frac{\tfu}\lu = \left[ 
\frac{(4\pi)^{n_\ssm + 2\du - 3}}{\lambda \sqrt{\pi \gtot/8}}
\frac{\lu}{\mpl}
\left(\frac{\mc}{\lu}\right)^{2k}
\right]^{1/(2\dsm + 2\du - 9)} < 1
\label{tfu}
\eeq

\begin{figure}[t]
\vspace{2.5in}
\centering
\includegraphics[bb=100 0 300 200,width=6cm]{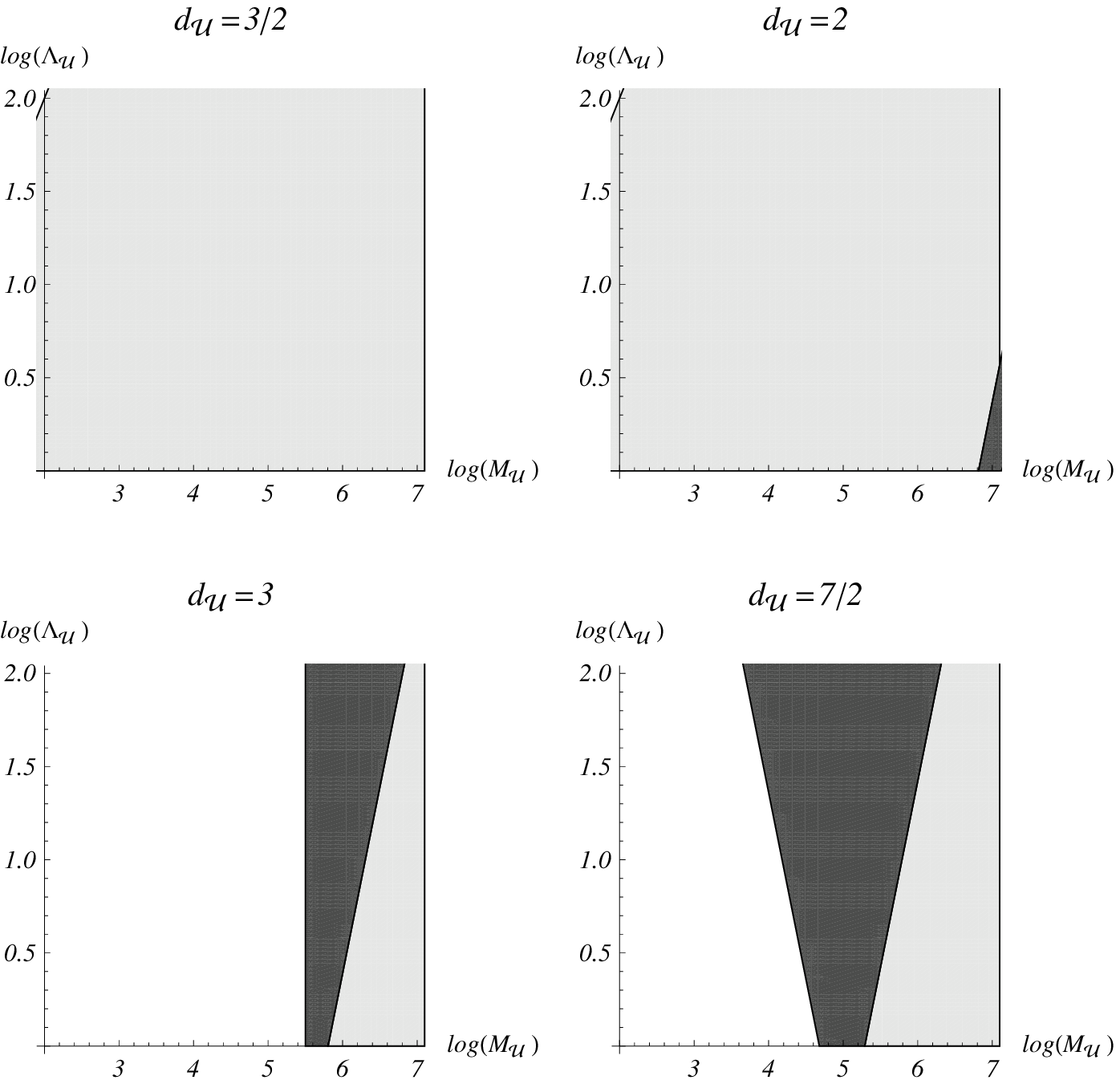}
%
%
\caption{Regions in the $\lu - \mc$ plane corresponding to various
freeze-out and thaw-in scenarios for $\du=3/2,~2,~3,~7/2$.
Dark grey:  SM-NP decoupling in the 
unparticle phase only; light grey: no SM-NP decoupling; in the white
regions $\tfu < v$ ($\lu, ~\mc$ are in TeV units).
We assumed $\gsm=\gbz=\gu=100$, $\gsm'=4$, $\gbz'=50$ and $\gu'=\du$.
For the $\bz$ phase: $n_{SM}=n_{NP}=2$, $\dsm=2$ and $d_{NP}=3$, while for
the $ \ucal $ phase:  $n_{SM}=2$, $n_{NP}=2(\du-1)$, $\dsm=2$ 
and $d_{NP}=\du$.}
\label{regions}
\end{figure}


For $ \du < 4.5 - \dsm $,
$ \Gamma/H $ has the singular property of
{\em increasing} as $T$ drops, whence SM
and NP will equilibrate for $ T < \tfu$ (thaw-in);
due to the constraints~\footnote{
The bounds on $ \du $ strictly hold in the conformal limit;
we expect deviations $ \propto g(T) - g_* 
\sim  (T/\lu)^a $ which we neglect.} on $ \du $ 
($\du < 1$ is excluded~ \cite{Grinstein:2008qk}) 
this can only happen for 
$ \ocal_\ssm = \phi^\dagger\phi $.
The opposite occurs if 
$ \du > 4.5 - \dsm $ (freeze-out). For $ \du = 4.5 - \dsm $, 
the approximations (\ref{eq:tot.en.de}), (\ref{gam}) are insufficient and
a detailed calculation is  required to determine 
freeze-out and/or thaw-in conditions; we will not consider this
special case further.

There are various possible scenarios for decoupling of the NP
sector.  The situation in the very early Universe ($T > \mc$)
depends on the UV completion (including the mediator interactions) of
the NP and will not be considered here.
If (\ref{tfbz}) holds then we have a standard freeze-out scenario:
the SM and NP sectors will be in equilibrium down to $T \sim \tfbz$
and decouple below this
value; thereafter the two sectors evolve keeping their entropies
separately conserved.
Since no mass thresholds or phase transitions are crossed~\footnote{We neglect
the possibility of right-handed neutrino decoupling.}
the SM and NP temperatures remain equal down to $ T \sim \lu $.

The situation for $\lu \gesim T $ is more complicated. If (\ref{tfu})
holds (which defines a region in the $ \lu-\mc$ plane), decoupling
occurs in the unparticle phase.  For $ T > v$ the most relevant
operator is $ \ocal_\ssm = \phi^\dagger\phi $, and both thaw-in (for
$\du < 2.5$) and freeze-out (for $\du > 2.5$) may be present.  For $ v
> T $ all the relevant SM operators have $\dsm=4$, and only freeze-out
is possible; in this case $ \tfu $ may be significantly smaller than
$v$.

Other parameter values lead to more complicated scenarios, e.g. a
double decoupling: freeze-out in the $\bz$, thaw-in in the unparticle
phase and then freeze out below $v$.  In spite of the many
possibilities, there is always a temperature below which the SM and NP
decouple.
  
In Fig.~\ref{regions} we show regions in the $(\lu,\mc)$ space that
correspond to various freeze-out and thaw-in scenarios 
for a reasonable parameter choice. For this calculation we
assumed that $\ocal_{\rm SM} = \phi^\dagger\phi$ is responsible for
maintaining the equilibrium between the SM and NP (so $\dsm=2$). For
consistency that choice implied an additional constraint $\tfu > v$
(below $v$ other SM operators are relevant). For interactions with the
$\bz$ phase an operator $\propto (\phi^\dagger\phi)(\bar
q_{\bz}q_{\bz})$, was adopted (in which case $\dbz=3$).  


\section{Big Bang Nucleosynthesis}
\label{bbn}

The light-element abundances resulting from BBN are sensitive to the
expansion rate that determines the temperature of the universe (see
e.g. \cite{Iocco:2008va}), which can be used to restrict possible
additional RDF, or, in our case, $ \gir$.  We express our results in
terms of the number of extra neutrino species, $ \Delta N_\nu$,
defined through 
\beq
\rnp =\frac{3}{\pi^2} \frac74 \left(\frac{4}{11}\right)^{4/3}
\Delta N_\nu T_\gamma^4 \,,
\label{delN}
\eeq
which is valid for $T$ below the $e^+e^-$ annihilation 
($T_\gamma$ stands for the photon temperature).  
For $ \Delta N_\nu$ we adopt the 
recent bounds obtained in
\cite{Iocco:2008va}:
$\Delta N_\nu = 0.0 \pm 0.3_{\rm stat}(2 \sigma) \pm 0.3_{\rm syst} $.

We first consider the case where SM and NP 
were in equilibrium down to a temperature $ T_f > v$, and decoupled
thereafter. 
Then the entropy conservation for the NP and SM sectors implies
\bea
\gnp^\star(T_f) (T_f R_f)^3 &=& \gnp^\star(\tnp) (\tnp R)^3 \cr
\gsm^\star(T_f) (T_f R_f)^3 &=& \gsm^\star(T_\gamma) (T_\gamma R)^3
\label{entcon}
\eea
where $R_f$ is the scale factor at the decoupling while $R$
corresponds to temperature of photons $T_\gamma$ ($\tnp$ is the
corresponding NP temperature); $\gnp^\star$ and $\gsm^\star$ stand for
the NP and SM effective numbers of RDF conventionally~\cite{Kolb}
adopted for the entropy density. 
After $e^+e^-$ annihilation
neutrinos and photons generate the dominant SM contribution, but their
temperatures differ. Using standard expressions~\cite{Kolb} we find
\beq 
\gsm^\star(T_\gamma) = g_\gamma \frac{g_\gamma + g_e + g_\nu}{g_\gamma+g_e}\,,
\label{gsm}
\eeq
where $g_i$ stands for the number of RDF corresponding to the species
$i$.  Assuming that $\gnp$ is almost constant in the temperature range
we are interested in and neglecting possible right-handed neutrino
decoupling effects, the two sectors had the same temperature down to
the electroweak phase transition; thereafter the temperatures split as
the SM crossed its various mass thresholds and the entropy was pumped
into remaining species.  Entropy conservation (\ref{entcon}) in both
sectors then implies
\beq
\tnp=T_\gamma\left[\frac{g_\gamma}{g_\gamma+g_e} \frac{g(\gamma,e,\nu)}{\gsm(v)}\right]^{1/3}
\label{tnp}
\eeq
where
$g_\ssm(\gamma, e,\nu)\equiv g_\gamma+g_e+g_\nu$,
while $\gsm(v)$ stands for the total number of SM RDF active above $T=v$. Note that the above
relation holds regardless if the decoupling happened during the $\bz$ or unparticle phase. 
Then combining with (\ref{delN}) we obtain
\beq
\gir = \frac74 \left[ \frac{\gsm(v)}{\gsm(\gamma,e,\nu)} \right]^{4/3} \Delta N_\nu
\label{gir}
\eeq
Using the standard expressions
for the SM quantities~\cite{Kolb} the
BBN constraint on $\Delta N_\nu$ then implies $\gir \lsim 20$ at 95\% CL.
It is worth mentioning here that  $ \Gamma $ measures
the decay rate of unparticles into SM states. After decoupling, when $ \Gamma < H $
these decays become very rare (the $\snp \to \ssm$ life-time becomes larger
than the age of the universe $ \sim1/H$).
   
More severe constraints could be obtained if NP and SM remained in equilibrium down
to the BBN temperature. That occurs for $\lu,~\mc \sim \tev$ and $
\du \sim 1 $; the relevant operator being $B_{\mu\nu} B^{\mu\nu}
\ocal_\ucal$.  Then, since temperatures of the NP and SM sectors are the same,
one obtains 
\beq
\gir = \frac74 \left(\frac{g_\gamma}{g_\gamma+g_e}\right)^{4/3} \Delta N_\nu
\eeq
which leads to $\gir \lsim 0.25$ at 95\% CL.

When decoupling occurs between $v$ and $T_{\rm BBN}$ the bound on $\gir $ 
lies between $0.25 $ and $ 20 $.
When
the SM and NP are never in equilibrium the BBN constraints can be used
to bound $ \rnp$, but not $ \gir $ since $\tnp $ is then not known.
  These bounds should be compared to 
$ \gir \gsim 100 $ typical of specific
models \cite{Banks:1981nn} {\it e.g.} for an
$\su3$ gauge theory with $16$ fundamental fermion multiplets, and 
expected from AdS/CFT correspondence \cite{Gubser:1999vj}. 
We conclude that many unparticle models will have difficulties accounting
for the observed light-element abundances.


\section{Summary}
Using the  trace anomaly
we argue for a form of the equation of state for unparticles
that contains power-like corrections to the 
expression for
relativistic matter; this allows us to determine temperature
dependence of the energy and entropy density for unparticles.
We then derive the Boltzmann equation for the $\bz$ phase and 
postulate
a plausible form for this equation for unparticles;
using this we determine the
conditions for NP-SM equilibrium. Finally we derive
useful constrains on the NP effective
number of degrees of freedom imposed
by the BBN.


\vspace{.5in}
\acknowledgments

This work was supported in part by the Ministry of Science and Higher
Education (Poland) as research projects N202~176~31/3844 (2006-8) 
and N~N202~006334 (2008-11) and by the U.S. Department of Energy
grant No.~DEFG03-94ER40837;
J.W.  was also supported in part by MICINN under contract SAB2006-0173.
B.G. acknowledges support of the
European Community within the Marie Curie Research \&\ Training
Networks:``HEPTOOLS" (MRTN-CT-2006-035505), and ``UniverseNet"
(MRTN-CT-2006-035863), and through the Marie Curie Host Fellowships
for the Transfer of Knowledge Project MTKD-CT-2005-029466.
J.W. acknowledges the support of the
MICINN project FPA2006-05294 and Junta de Andaluc{\'\i}a 
projects FQM 101, FQM 437 and FQM03048.


\appendix

\section{Derivation of the reaction rate using the Kubo formalism}
\label{sec:app.kubo}

In this section we follow closely the arguments presented in \cite{Kubo:1957mj}. We
consider a thermodynamic system, not necessarily in equilibrium, with macroscopic observables 
$ \{ \alpha_i\}$  associated with operators $ \{ a_i \}$. We assume the
thermodynamics of the system is described by a density matrix $ \rho $
\beq
\rho = \exp \left[ \beta \left( \Omega - H + \sum_i \mu_i a_i \right) \right]
\label{eq:den.ma}
\eeq
where the $ \mu_i $ and $ \beta $ are parameters, and $ \Omega  = \Omega( \mu,\beta)$
is a function chosen such that
tr$\rho=1 $, that is
\beq
e^{ - \beta \Omega } = \tr e^{ - \beta \left( H - \sum_i \mu_i a_i \right) }
\eeq
The $ \mu_i $ are determined by the condition
\beq
\alpha_i = \tr \rho a_i  = - \deriva\Omega{\mu_i}{}
 \label{eq:al.mu}
\eeq

It is important to note that  $ \rho $
differs from the usual  grand-canonical
density operator in that the  $a_i$ are not assumed 
to be conserved, so the $ \alpha_i $ will not be constant:
\beq
\alpha_i(t) = \tr \{ \rho \; a_i(t) \} = \tr \{ \rho(t) \; a_i \}; \qquad
a_i(t) = e^{ i H t } a_i e^{-i H t}, ~~
\rho(t) = e^{ - i H t } \rho e^{i H t}
\eeq
$ \alpha_i(t) $  denotes the average of $a_i $
at time $t$ for a distribution for which the average of $a_i $ at 
$ t=0 $ is $ \alpha_i  = \alpha_i(0)$.

We now assume the $ \mu_i $ are small,
then a straightforward calculation yields
\beq
\Omega = \Omega_0 - \sum_i \mu_i \vevof{a_i} + \cdots \,,
\eeq
where, for any operator $ \xi $,
\beq
\vevof\xi = \tr \rho_0 \xi; \qquad \rho_0 = e^{ \beta( \Omega_0 -H)} ,
~~ e^{ - \beta \Omega_0 } = \tr e^{ - \beta H} \,.
\label{eq:av.def}
\eeq

Now let
\beq
 \alpha'_i(t) 
 = \tr \rho a'_i(t) 
= \tr \left\{ \rho e^{ i H t } a_i' e^{ - i H t } \right\} \,;
\quad a_i' = a_i - \vevof{a_i} \,,
\eeq
so that, to first order in $ \mu $,
\beq
\alpha'_i(t) =  \sum_j
\int_0^\beta ds \vevof{ a'_j(- i s) a'_i(t) } \mu_j
\label{eq:alipt}
\eeq
Using now the cyclic property of the trace, $
\vevof{ \xi(z) \eta(z') } = \vevof{\xi(z-z')\; \eta}
 = \vevof{\xi\; \eta(z'-z)} $
for any operators $\xi,~\eta $
and any complex times $z,~z'$. From this it follows that
\beq
\frac{d^2}{d t^2} \vevof{a'_i(-i s) a'_j(t)}
 = - \vevof{\dot a_i'(-is) \dot a'_j(t) } \,,
\eeq
hence
\beq
\int_0^\tau dt \left( 1 - \frac t\tau \right)
\vevof{\dot a_i'(-is) \dot a'_j(t) }
= \vevof{a_i'(-is) \dot a'_j(0) }
- \inv\tau \left[
\vevof{a_i'(-is) a'_j(\tau) } -
\vevof{a_i'(-is) a'_j } \right] \,.
\eeq
Next, using the definition
\beq
\dot \xi(t) = i \left[ H , \xi(t) \right]
\eeq
and the cyclic property of the trace,
\beq
\int_0^\beta ds \vevof{a_i'(-is) \dot a'_j }
= - i \vevof{[a_i' , a_j' ]}
= - i \vevof{[a_i , a_j ]} \,.
\eeq
Collecting all results and using $ \dot a_i' = \dot a_i $,
\bea
\frac{\alpha'_i(\tau) - \alpha'_i(0)}\tau
&=& - \sum_j \gcal(\tau)_{ij} \mu_j \cr
\gcal(\tau)_{ij} &=&
\int_0^\beta ds \int_0^\tau dt \left( 1 - \frac t\tau \right)
\vevof{ \dot a_j(-i s) \dot a_i(t) } + i \vevof{ [a_i, a_j] }
\cr &&
\label{eq:Kubo}
\eea
which is the celebrated Kubo equation.
It is important to note that the
$ \tau \to 0 $ limit is subtle \cite{Kubo:1957mj}.

Suppose that the system is composed
of two sub-systems, labeled `$1$' and `$2$' with a
Hamiltonian
\beq
H = H_1 + H_2 + \epsilon H' \, ; \qquad  [ H_1 , H_2] =0 \,,
~~ \epsilon \ll 1
\label{eq:ham}
\eeq
and take $ a_1 = H_1 \,, ~ a_2 = H_2$; in this case $ \rho $
describes two systems at different temperatures
that weakly interact through $ \epsilon H' $. Then
\beq
\alpha_i = \vevof{H_i} = V \rho_i 
\eeq
where $ \rho_i $ denotes the energy density
and $V$ the space volume of the system. We imagine that 
each subsystem has a well defined temperature $T_i $
but that these change slowly due to the presence of $ H' $;
we also require the systems to be close to equilibrium with
each other so that $ | T - T_i| \ll T $. In this case
the left hand side of (\ref{eq:Kubo}) corresponds to 
$ \dot\alpha_i' $ while on the right hand side we
can take the $ \tau \to \infty $ limit 
since the
integrand is damped at times larger than the
characteristic times of systems $1$ and $2$;
see Ref.~\cite{Kubo:1957mj} for details. In this case
\beq
\dot \rho_i = c_i \dot {\delta T}_i \,; \quad \delta T_i = T_i - T
\eeq
where $c_i $ denote the heat capacities per unit volume at temperature
$T$.

When $ \epsilon  =0 $, the density matrix (\ref{eq:den.ma}) becomes
\beq
\left. \rho \right|_{\epsilon  =0 }
= e^{ \beta \Omega - \beta ( 1 - \mu_1 ) H_1 - \beta ( 1 - \mu_2 ) H_2}
\eeq
which corresponds to non-interacting subsystems at temperatures
$ T_i =  T/(1-\mu_i) $, whence
\beq
\mu_i = \inv T \delta T_i \,, \quad T = \inv\beta; \qquad ( \epsilon =0 )
\label{eq:mu.dt}
\eeq
Then (\ref{eq:Kubo}) gives
\beq
V c_i \dot{\delta T}_i = - \inv T \sum_j \gcal_{i j} \; \delta T_j \,;
\qquad
\gcal_{ij} =
\int_0^\beta ds \int_0^\infty dt 
\vevof{ \dot H_j(-i s) \dot H_i(t) } 
\label{eq:evol}
\eeq
where
\beq
\dot H_i = i [ H , H_i] = i \epsilon [ H' , H_i]
\then \dot H_i(z) = e^{ i z H} \dot H_i e^{ - i z H} = O(\epsilon ) \,,
\eeq
so that $ \gcal $ is of order
$ \epsilon ^2 $; since we work to the lowest non-trivial order in 
$H'$, this also justifies the use of (\ref{eq:mu.dt}).

Now we need to evaluate $ \gcal $. Following (\ref{eq:Lint}), we assume
\beq
H' = - \int d^3\xx \ocal_1 \ocal_2
\eeq 
then
\beq
i [ H' , H_1]_{\epsilon =0 }
=   \int d^3\xx \; i [H_1, \ocal_1] \ocal_2 
=  \int d^3\xx \dot \ocal_1 \ocal_2
\eeq
and, similarly, $ i [ H , H_2] = \int d^3\xx \dot \ocal_2 \ocal_1 $
>From this
\bea
\left\{ \inv{\epsilon^2}\vevof{\dot H_1(-is) \dot H_1(t)}
\right\}_{\epsilon =0 } 
= \int d^3\xx\; d^3\yy \vevof{ \dot\ocal_1(-is,\xx)
\dot\ocal_1(t,\yy)} \vevof{\ocal_2(-is,\xx) \ocal_2(t,\yy)} \cr
\left\{ \inv{\epsilon^2}\vevof{\dot H_1(-is) \dot H_2(t)}
\right\}_{\epsilon =0 } 
= \int d^3\xx\; d^3\yy \vevof{ \dot\ocal_1(-is,\xx)
\ocal_1(t,\yy)} \vevof{\ocal_2(-is,\xx) \dot \ocal_2(t,\yy)} \cr
\left\{ \inv{\epsilon^2}\vevof{\dot H_2(-is) \dot H_1(t)}
\right\}_{\epsilon =0 } 
= \int d^3\xx\; d^3\yy \vevof{ \ocal_1(-is,\xx)
\dot\ocal_1(t,\yy)} \vevof{\dot \ocal_2(-is,\xx) \ocal_2(t,\yy)} \cr
\left\{ \inv{\epsilon^2}\vevof{\dot H_2(-is) \dot H_2(t)}
\right\}_{\epsilon =0 } 
= \int d^3\xx\; d^3\yy \vevof{ \ocal_1(-is,\xx)
\ocal_1(t,\yy)} \vevof{\dot\ocal_2(-is,\xx) \dot\ocal_2(t,\yy)}
\eea
where the $\vevof{\cdots}$ separates into a product because
when $ \epsilon =0 $ averages separate into averages over
systems $1$ and $2$ which are independent.
For the case where the $ \ocal_i $ are scalars and even under
time reversal all the above correlators are equal up to a sign,
so that
\beq
\gcal = \epsilon^2 G V \bpm{ 1 &-1 \cr -1 & 1} \epm
\label{eq:G.simple}
\eeq
where $V$ denotes the volume of space and
\beq 
G = \int_0^\beta ds \int_0^\infty dt \int d^3\xx\;
\vevof{\ocal_1(-i s , \xx) \dot\ocal_1(t,{\bf0})}
\vevof{\ocal_2(-i s , \xx) \dot\ocal_2(t,{\bf0})}
\eeq

Substituting (\ref{eq:G.simple}) in (\ref{eq:evol}) gives
$  c_1 \dot{\delta T}_1 = 
- c_2 \dot{\delta T}_2 = - (\epsilon^2 G/T) (\delta T_1 - \delta T_2) $,
then
\beq
\partial_t ( {\delta T}_1 - {\delta T}_2) = 
- \Gamma ( {\delta T}_1 - {\delta T}_2) \,; \quad
\Gamma =
\left( \inv{c_1} + \inv{c_2} \right) 
\frac{ (\epsilon^2 G) }T
\label{eq:gamma.kubo}
\eeq

The quantity $G$ can be evaluated using the tools of 
finite-temperature field theory.
To facilitate this let
\beq
J_0 = - i \ocal_1 \stackrel\leftrightarrow{\partial_t}
\ocal_2
\label{eq:def.of.j0}
\eeq
then, setting $ \epsilon =0 $
and using invariance under space translations,
\bea
G &=& -\inv4 \int_0^\beta ds \int_0^\infty dt \int d^3\xx
\vevof{J_0(-i s , \xx) J_0(t,{\bf0})} \cr
&=& - \inv4 \Re\left\{  \lim_{\omega,\kk \to 0}  \int_0^\beta ds
\int d^3\xx\; dt \; e^{ -i ( \omega t - \kk \cdot \xx)} \theta(t)
\vevof{J_0(-i s , {\bf0}) J_0(t,\xx)} \right\}
\eea
In this form $G$ can be evaluated in terms of the
correlator of two $J_0$ currents.
We took the real part, which is the one that yields the
relevant width, and introduced $k$ as a regulating 4-momentum.
The limit $\omega,~\kk \to 0 $ requires care, for the present 
case one should first set $ \kk =0 $ and then take
$ \omega $ to zero~\cite{mahan}. 

In order to  compare
this result to the one derived using the Boltzmann  equation
it proves convenient to do a Lehmann expansion of $G$, which
involves matrix elements of the form
$ \vevof{n | J_0 |m } $. In terms of Feynman graphs, such 
matrix elements will include pieces that are 
not connected to $J_0 $; these disconnected pieces factorize
and cancel the factor $ \exp(\beta \Omega_0 ) $~\cite{mahan}
that appears in the definition of the average (\ref{eq:av.def}).
We then find
\beq
G = \inv8 \sum e^{ \beta( \Omega_0 -  E_n )} \beta 
\left| \vevof{ n | J_0 | m }_{\rm con}
\right|^2 (2 \pi)^4 \delta\up4( p_n - p_m ) 
\label{eq:lehmann}
\eeq

\bigskip

Up to now we have assumed that the volume of the system is kept fixed,
but this can be easily relaxed. The calculation involves obtaining
the thermodynamic potential to order $ \mu ^2$ and will not be 
presented here, the final result is the expected one:
the time evolution equation becomes $ \dot \rho_i + 4 H \rho_i 
 = - \sum \gcal_{ij} \delta T_j/T $ where $ \dot V/V = - 3 H $.

\section{The Boltzmann equation}
\label{sec:app.boltzmann}

We again imagine two sectors, labeled 1 and 2; within each 
the interactions are strong enough to maintain equilibrium
at temperatures $T_{1,2} $; the sectors interact
only though (\ref{eq:Lint}). We denote by
$ \nu_{a\up i} $  the
distributions of particles $a$ in sector $i$; the
corresponding Boltzmann equation is
\beq
p^\alpha \deriva{\nu_{a\up i}}{x^\alpha}{}
- \Gamma^\alpha_{\beta \gamma} p^\beta p^\gamma 
\deriva{\nu_{a\up i}}{p^\alpha}{}
= \CC[\nu_{a\up i}]
\label{eq:gen.BE}
\eeq
where the \rhs\ denotes the collision term. 

We consider first a process of the form 
$ X_1 + X_2 \to X_1' + X_2' $,
where $X_i,~X_i'~~ (i=1,2) $ denote states in system $i$.
If a particle labeled by $ a\up1$ is in $X_1$, then the
corresponding collision term $\CC[\nu_{a\up1}]$ is given by
\bea
\CC_{X,X'}[\nu_{a\up1}]
&=& - \int d\Phi'_{X,X'} \half \left| \mcal( X \to X' ) \right|^2
(2 \pi)^4 \delta( K_1 + K_2 - K_1' - K_2') 
\ncal_{X,X'} \cr
\ncal_{X , X'} &=&
\left[
\prod_{b\up1 \in X_1'} \left( 1 \pm \nu_{b\up1} \right)
\prod_{c\up2 \in X_2'} \left( 1 \pm \nu_{c\up2} \right)
\prod_{d\up1 \in X_1} \nu_{d\up1} 
\prod_{e\up2 \in X_2} \nu_{e\up1} \right]
- \cr && \qquad \qquad 
\left[
\prod_{b\up1 \in X_1} \left( 1 \pm \nu_{b\up1} \right)
\prod_{c\up2 \in X_2} \left( 1 \pm \nu_{c\up2} \right)
\prod_{d\up1 \in X_1'} \nu_{d\up1} 
\prod_{e\up2 \in X_2'} \nu_{e\up1} \right]
 \cr \cr
K_i &=& ( E_i, \KK_i) , \quad
E_i = \sum_{a\up i \in X_i} E_{a\up i} \qquad
\KK_i = \sum_{a\up i \in X_i} \kk_{a\up i} \cr
K_i' &=& ( E_i', \KK_i') , \quad
E_i' = \sum_{a\up i \in X_i'} E_{a\up i} \qquad
\KK_i' = \sum_{a\up i \in X_i'} \kk_{a\up i}
\eea
where $ d\Phi'_{X,X'} $ denotes the corresponding invariant
phase space measure for all particles except $a\up1$ (as indicated 
by the prime), $ \mcal $ the Lorentz-invariant matrix element,
and $E_a,~ \kk_a$ denote  the energy an momentum of particle $a$.
The upper sign corresponds to bosons, the lower to fermions.

We will assume spatial homogeneity, so that the $ \nu $ will
depend only on time and energy, and also
assume kinetic equilibrium, so that the density
functions take the usual  Fermi-Dirac or Bose-Einstein form, but
with time dependent temperature and, possibly, chemical potential.
Then
\bea
\ncal_{X,X'} &=& 
\left( e^{- E_1/T_1 - E_2/T_2 } - 
e^{- E_1'/T_1 - E_2'/T_2 } \right) \nu_{X ,X'};\cr
\nu_{X , X'} &=&
\prod_{b\up2 \in X_2,X_2'} \left( 1 \pm \nu_{b\up2} \right)
\prod_{c\up1 \in X_1,X_1'} \left( 1 \pm \nu_{c\up1} \right);
\eea

Using this we can derive the time dependence of the energy
density; for simplicity we will carry out the calculation in flat space.
The energy density associated with the $ a\up1 $ is
\beq
\rho_{a\up1} = \int \frac{d^3\pp}{(2\pi)^3} E_{a\up1} \nu_{a\up1}
= 2 \int d \Phi_{a\up1} E_{a\up1}^2  \nu_{a\up1}
\eeq
Integrating (\ref{eq:gen.BE}) over \pp\ we find
\beq
\left. \partial_t \rho_{a\up1} \right|_{X,X'} = 
- \int 
d\Phi_{X,X'} E_{a\up1} \left| \mcal( X \to X' ) \right|^2
(2 \pi)^4 \delta( K_1 + K_2 - K_1' - K_2' ) 
\ncal_{X,X'} \,, \quad
d \Phi_{X,X'} = d\Phi_{X,X'}' d\Phi_{a\up1} \,;
\eeq
where the notation on the \lhs\ indicates that this corresponds to the
change in $ \rho_{a\up1} $ generated by this particular $X \to X' $ reaction.
The total time derivative is obtained by summing over
all states $X,X'$ such that $a\up1 \in X_1$:
\beq
\dot \rho_{a\up1} = 
- \sum_{X,X';\; (a\up1 \in X_1)}\int 
d\Phi_{X,X'} E_{a\up1} \left| \mcal( X\to X' ) \right|^2
(2 \pi)^4 \delta( K_1 + K_2 - K_1' - K_2' ) 
\ncal_{X,X'} 
\eeq
The time derivative of the total energy
density for each sector is then obtained by now summing over all $a\up1$:
\beq
\dot \rho_1 = 
- \sum_{X,X'}\int 
d\Phi_{X,X'} E_1 \left| \mcal( X \to X' ) \right|^2
(2 \pi)^4 \delta( K_1 + K_2 - K_1' - K_2' ) 
\ncal_{X,X'} 
\eeq

To make this look more symmetric consider the contribution
with $X$ and $X'$ exchanged. Since $ |\mcal|^2 $ is the same
but $ \ncal $ changes sign we can write
\beq
\dot \rho_1 = 
-  \half\sum_{X,X'}\int 
d\Phi_{X,X'} (E_1 - E_1') \left| \mcal( X \to X' ) \right|^2
(2 \pi)^4 \delta( K_1 + K_2 - K_1' - K_2' ) 
\ncal_{X,X'} 
\eeq
The corresponding expression for  $ \dot \rho_2 $
is obtained by switching the $1$ and $2$ subscripts.

We are interested in cases where the Maxwell-Boltzmann statistics
are adequate, so $ \nu_{X,X' } \simeq 1 $, and when the 
temperatures are similar: $ T_i = T + \delta T_i$.
Using the  energy conservation condition $ E_1 + E_2
= E_1' + E_2' $, we find
\beq
\ncal_{X,X'} \simeq - e^{-(E_1+E_2)/T} \frac{E_1 - E_1'}{T^2}
\left( \delta T_1 - \delta T_2 \right)
\eeq
Also, ignoring non-relativistic contributions to the energy
density
\beq
\dot \rho_i = c_i \dot \delta T_i
\eeq
where $ c_i $ is the heat capacity per unit volume. Collecting
all expressions gives
\bea
&& \partial_t \left( \delta T_1 - \delta T_2 \right) = - \Gamma
\left( \delta T_1 - \delta T_2 \right) \,, \cr \cr
&& \Gamma = \left( \inv{c_1} + \inv{c_2} \right)
\inv{2 T}  \sum_{X',X}
\int d\Phi_{X,X'} \beta (E_1 - E_1')^2 e^{ - \beta E_X}
\left| \mcal( X \to X' )\right|^2 (2\pi)^4 \delta (K_X - K_{X'}) \,.
\eea

In order to compare this with the Kubo formula we use
\bea
 \mcal( X \to X' ) 
&=& \vevof{ X' | \lcal_{\rm int} | X} 
= \epsilon \vevof{ X' |\ocal_1 \ocal_2 | X}
\eea
where we work to lowest non-trivial order in the interaction.
Using $J_0$, defined in (\ref{eq:def.of.j0}), we find
\beq
\vevof{X' |J_0| X}_{\epsilon =0 } =
2 (E_1 - E_1') \vevof{ X' | \ocal_1\ocal_2 | X}_{\epsilon =0 }
\eeq
where we took $ \epsilon =0 $ since we are interested
only in the leading contributions to $ \Gamma $. Then
\beq
\Gamma = \left( \inv{c_1} + \inv{c_2} \right)
\frac{\beta^2 \epsilon^2}8  \sum_{X',X}
\int d\Phi_{X,X'} e^{ - \beta E_X}
\left|\vevof{X' |J_0| X}\right|^2 (2\pi)^4 \delta (K_X - K_{X'})
\eeq
Using then the Lehmann expansion (\ref{eq:lehmann}) we find
\beq
\Gamma = \left( \inv{c_1} + \inv{c_2} \right)
 \frac{\epsilon^2 |G|}T 
\eeq
exactly as in the Kubo formalism~\footnote{We have used the 
Boltzmann approximation in identifying $E_n$ in (\ref{eq:lehmann}), which is
the total energy of state $\ket n$, with $E_X$ which is the sum
of the energies of the particles in state $ \ket X $. These 
energies are
approximately equal for a sparse system, where this approximation holds.}.

\bigskip

Despite  its intuitive appeal the Boltzmann approach contains
conceptual difficulties for the case of strongly interacting
theories, for which concepts such as the particle densities
$ \nu_a $ are ill defined. In this case the definition of
$ \Gamma $ (\ref{eq:gamma.kubo})
obtained through the Kubo equation is preferable
where the relevant matrix elements can, at least in principle, be
obtained numerically.



\begin{thebibliography}{99}

  
  
\bibitem{Georgi:2007ek}
  H.~Georgi,
  Phys.\ Rev.\ Lett.\  {\bf 98}, 221601 (2007)

\bibitem{Georgi:2007si}
  H.~Georgi,
  Phys.\ Lett.\  B {\bf 650}, 275 (2007)
  
\bibitem{van der Bij:2006pg}
  J.~J.~van der Bij and S.~Dilcher,
  Phys.\ Lett.\  B {\bf 638}, 234 (2006)
  
\bibitem{Banks:1981nn}
  T.~Banks and A.~Zaks,
  Nucl.\ Phys.\  B {\bf 196}, 189 (1982).

\bibitem{Davoudiasl:2007jr}
  H.~Davoudiasl,
  arXiv:0705.3636 [hep-ph].
  
\bibitem{McDonald:2007bt}
  J.~McDonald,
  arXiv:0709.2350 [hep-ph].
  
\bibitem{Lewis:2007ss}
  I.~Lewis,
  arXiv:0710.4147 [hep-ph].
 
\bibitem{Chen:2007qc}
  S.~L.~Chen, X.~G.~He, X.~P.~Hu and Y.~Liao,
  arXiv:0710.5129 [hep-ph].
   
\bibitem{Kikuchi:2007az}
  T.~Kikuchi and N.~Okada,
  arXiv:0711.1506 [hep-ph].

\bibitem{cft.temp} See, e.g.,
  J.~M.~Maldacena,
  arXiv:hep-th/0309246;
%
  O.~Aharony, S.~S.~Gubser, J.~M.~Maldacena, H.~Ooguri and Y.~Oz,
  Phys.\ Rept.\  {\bf 323}, 183 (2000)
and references therein.

\bibitem{Collins:1976yq}
  J.~C.~Collins, A.~Duncan and S.~D.~Joglekar,
  Phys.\ Rev.\  D {\bf 16}, 438 (1977).
  
\bibitem{Gubser:1999vj}
  S.~S.~Gubser,
  Phys.\ Rev.\  D {\bf 63}, 084017 (2001)

\bibitem{Svetitsky:2009pz}
  B.~Svetitsky,
  arXiv:0901.2103 [hep-lat].

\bibitem{Kolb}
  E.~W.~Kolb and M.~S.~Turner,
{\it Addison-Wesley (1990)}
  
\bibitem{Stephanov:2007ry}
  M.~A.~Stephanov,
  Phys.\ Rev.\  D {\bf 76}, 035008 (2007)

\bibitem{Rajaraman:2008bc}
  A.~Rajaraman,
  arXiv:0806.1533 [hep-ph].
  

\bibitem{Grinstein:2008qk}
  B.~Grinstein, K.~Intriligator and I.~Z.~Rothstein,
  Phys.\ Lett.\  B {\bf 662}, 367 (2008)
  G.~Mack,
  Commun.\ Math.\ Phys.\  {\bf 55}, 1 (1977).
  
\bibitem{Iocco:2008va}
  F.~Iocco, G.~Mangano, G.~Miele, O.~Pisanti and P.~D.~Serpico,
  arXiv:0809.0631 [astro-ph].

\bibitem{Kubo:1957mj}
  R.~Kubo,
  J.\ Phys.\ Soc.\ Jap.\  {\bf 12}, 570 (1957).
%
  R.~Kubo, M. Yokota and S. Kakajima,
  J.\ Phys.\ Soc.\ Jap.\  {\bf 12}, 1203 (1957).

\bibitem{bern}
J. Bernstein, {\it Kinetic theory in the expanding universe}
Cambridge monographs on mathematical physics,
(Cambridge University Press, New York, 1988). 


\bibitem{mahan}
G.~D.~Mahan, {\it Many-particle physics}
(Plenum, New York, 1990)

\end{thebibliography}
\end{document}